\documentclass[floatfix,aps,prl,onecolumn,superscriptaddress,10pt]{revtex4-2}

\usepackage{graphicx,xcolor}
\usepackage{epstopdf}
\usepackage{dcolumn}
\usepackage{bm}
\usepackage{bbold}
\usepackage{float}
\usepackage[UKenglish]{babel}
\usepackage[utf8]{inputenc} %adds non-UKc standard accents
\usepackage[centerlast,small]{caption}
\usepackage{subcaption}
\usepackage[normalem]{ulem}
\usepackage{tabularx}
\usepackage{hyperref}

\usepackage{amsmath, esint}
\usepackage[separate-uncertainty=true]{siunitx}
\DeclareSIUnit{\sqrthz}{\sqrt\mathrm{Hz}}

\makeatletter
\renewcommand{\fnum@figure}{ Fig. \thefigure}
\makeatother

\usepackage{titlesec}
\titleformat{\section}
{\normalfont\bfseries\centering}
{}{\thesection.}{}

\begin{document}
\title{Quantum magnetic imaging of current density in lithium-ion batteries}

% Blank author list
\author{W.\,Evans*}
\affiliation{Department of Biosignals, Physikalisch-Technische Bundesanstalt, 10587 Berlin, Germany}
\author{T.\,Coussens}
\affiliation{Department of Physics and Astronomy, University of Sussex, Brighton, BN1 9RH, United Kingdom}
\author{M.\,T.\,M.\,Woodley}
\affiliation{Department of Physics and Astronomy, University of Sussex, Brighton, BN1 9RH, United Kingdom}
\affiliation{Centre for Photonics, Department of Physics, University of Bath, BA2 7AY, United Kingdom}
\author{A.\,M.\,Fabricant}
\affiliation{Department of Biosignals, Physikalisch-Technische Bundesanstalt, 10587 Berlin, Germany}
\author{G.\,D.\,Kendall}
\affiliation{CDO² Germany, Salzdahlumer Straße 196, 38126 Braunschweig, Germany}
\author{M.\,Sonnet}
\affiliation{Chair for Electrochemical Energy Conversion and Storage Systems (ESS) - Institute for Power Electronics and Electrical Drives (ISEA), RWTH Aachen University, Campus-Boulevard 89, 52074 Aachen, Germany}
\affiliation{Center for Ageing, Reliability and Lifetime Prediction of Electrochemical and Power Electronic Systems (CARL), RWTH Aachen University, Campus-Boulevard 89, 52074 Aachen, Germany}
\author{D.\,Wasylowski}
\affiliation{Chair for Electrochemical Energy Conversion and Storage Systems (ESS) - Institute for Power Electronics and Electrical Drives (ISEA), RWTH Aachen University, Campus-Boulevard 89, 52074 Aachen, Germany}
\affiliation{Center for Ageing, Reliability and Lifetime Prediction of Electrochemical and Power Electronic Systems (CARL), RWTH Aachen University, Campus-Boulevard 89, 52074 Aachen, Germany}
\author{D.\,U.\,Sauer}
\affiliation{Chair for Electrochemical Energy Conversion and Storage Systems (ESS) - Institute for Power Electronics and Electrical Drives (ISEA), RWTH Aachen University, Campus-Boulevard 89, 52074 Aachen, Germany}
\affiliation{Center for Ageing, Reliability and Lifetime Prediction of Electrochemical and Power Electronic Systems (CARL), RWTH Aachen University, Campus-Boulevard 89, 52074 Aachen, Germany}
\affiliation{Helmholtz Institute Münster: Ionics in Energy Storage (HI MS), IMD-4, Forschungszentrum Jülich, 52425 Jülich, Germany}
\author{F.\,Oru\v{c}evi\'{c}}
\affiliation{Department of Physics and Astronomy, University of Sussex, Brighton, BN1 9RH, United Kingdom}
\author{P.\,Kr\"{u}ger}
\affiliation{Department of Biosignals, Physikalisch-Technische Bundesanstalt, 10587 Berlin, Germany}
\affiliation{Department of Physics and Astronomy, University of Sussex, Brighton, BN1 9RH, United Kingdom}
\affiliation{CDO² Germany, Salzdahlumer Straße 196, 38126 Braunschweig, Germany}

%----------------------- Abstract -------------------------
\begin{abstract}
The projected rapid growth of battery cell production over the next decade demands advanced diagnostic tools for quality control, ageing prediction, and recycling. Most existing techniques lack the spatial and temporal resolution required to capture internal electrochemical processes non-invasively. Here, we present magnetic imaging of current densities in battery cells, a sensitive quantum-magnetometry method that uses optically pumped magnetometers (OPMs) to perform real-time imaging of internal dynamics in open-circuit configuration. We demonstrate this approach for monitoring relaxation processes in 6000\,mA\,h lithium-ion cells following pulsed discharges across a range of pulse durations and currents as well as states of charge. The measurement results are benchmarked against superconducting-quantum-interference-device (SQUID) magnetometry and validated with three-dimensional finite element simulations. Equivalent circuit models are employed to interpret the relaxation profiles, revealing spatially resolved features and transient magnetic-field signatures that are inaccessible with complementary non-invasive techniques such as electrochemical impedance spectroscopy (EIS). This work establishes OPM-based magnetic imaging of battery current density as a powerful diagnostic tool with potential impact on cell development, manufacturing quality assurance, and second-life assessment.
\end{abstract}

%----------------------- Abstract -------------------------
\maketitle

%--------------------- Introduction -----------------------
% \section*{Introduction}
The success of the green-energy transition relies on scalable, high-performance batteries to power electric vehicles, integrate renewable-energy systems, and enable ambitious emerging technologies such as electric flight~\cite{Hecht2022}.
Yet, despite rapid advances in cell chemistry and manufacturing, a fundamental bottleneck remains: a lack of diagnostic tools that can non-invasively monitor internal electrochemical processes with both spatial and temporal resolution. 
Such tools are required to deepen our understanding of defect formation and ageing of currently used battery cells as well as those under development. 
Beyond the key fundamental research needs, present limitations in battery characterisation hinder the optimisation of manufacturing, shorten cell lifetimes, and lead to safety risks and costly inefficiencies. 
In ramp-up gigafactories, scrap rates can exceed 30\%~\cite{Naumann2023}, and the inability to detect inhomogeneities or defects during formation or operation directly impacts performance and reliability. 
As second-life applications expand, efficient triage of used cells requires advanced non-destructive methods for assessing local state of health (SoH)~\cite{Borner2022}.

In this work, we present a quantum-sensing approach that overcomes these limitations. 
Using optically pumped magnetometers (OPMs), also known as atomic magnetometers, we demonstrate non-invasive magnetic imaging for spatially resolving current density dynamics in lithium-ion cells. 
OPMs combine quantum-limited sensitivity, room-temperature operation, and scalability~\cite{Schwindt2007}, offering a practical and disruptive alternative to existing tools. 
Unlike superconducting quantum interference devices (SQUIDs)~\cite{Bechstein2020,Storm2016,Storm2019,Jaklevic1964}, which require cryogenics, or nitrogen-vacancy (NV) centers in diamond~\cite{Zhang2021,Hollendonner2023,Steinert2010,Pollok2025}, which face challenges with sensitivity and field of view, OPMs enable real-time magnetic imaging of larger-format cells under realistic operating conditions.

Conventional battery diagnostics focus on global parameters such as capacity fade~\cite{Lewerenz2020}, over-voltage hysteresis~\cite{Ovejas2019}, or ohmic resistance~\cite{Tomaszewska2019}, which provide only limited insights into internal dynamics. 
Complex cell constructions~\cite{Zhang2013,Zhang2023} can provide some access to internal processes, but are limited in their industrial application. 
Advanced imaging techniques---including nuclear magnetic resonance (NMR)~\cite{Pecher2017,Tucker2002,Romanenko2021}, electron microscopy~\cite{Li2017,Zachman2018}, and acoustic scanning~\cite{Wasylowski2022,Wasylowski2024}---are destructive or too slow for real-time monitoring. 
Electrochemical impedance spectroscopy (EIS)~\cite{Agarwal1992,Meddings2020,Jones2022}, though widely used, lacks the spatial resolution needed to detect localised phenomena. Magnetic imaging offers a unique path forward, providing a direct link between internal current densities and electrochemical activity throughout the battery’s charge-discharge cycle~\cite{Hunter2025}. 
Previous work has explored magnetic susceptibility~\cite{Hu2020,Hu2020a}, slow scanning~\cite{Suzuki2021}, and current mapping with classical magnetometers~\cite{Bason2022,Brauchle2023,Pang2013,Janosek2009,Butta2010}, but none of these methods provide the combination of sensitivity, spatial resolution, and real-time monitoring that quantum magnetometry can offer. This work demonstrates, for the first time, quantum magnetic imaging current densities in commercial lithium batteries, with the ability to track dynamic magnetic fields signals over 100 second duration down to 500 femtotesla.

To illustrate the method, we employ an OPM array to measure magnetic-field relaxation following pulsed 0.08--0.3\,C  discharges of 6000\,mA\,h lithium-ion cells.
This discharge rate was selected as a compromise between minimising heating effects and changes to state of charge (SoC), while optimising the signal-to-noise ratio of the relaxation signal. 
In this way, we monitor SoC-dependent relaxation dynamics, revealing significant spatial variations across the cell. 
Benchmark measurements using SQUID magnetometersconfirm the reliability and accuracy of the approach. We further combine three-dimensional finite element simulations~\cite{Doyle1994,Doyle1995} with equivalent circuit analysis to interpret the measured relaxation profiles~\cite{Bernardi2011,Roth2023}. 
Importantly, we observe localised transient magnetic-field signals, detectable with only a subset of sensors, which may be attributed to dynamic electrochemical events such as soft short-circuit formation.

Our results help establish quantum magnetic imaging as a breakthrough diagnostic tool for a deeper understanding of fundamental battery processes such as ageing. 
The method further bridges laboratory research and industrial battery manufacturing. 
By enabling real-time, spatially resolved characterisation of internal processes, this approach paves the way for more reliable cell designs, improved quality control in gigafactories, and efficient second-life assessment of batteries.
%--------------------- Introduction -----------------------

%---------------------- Experiment ------------------------
\section*{Methodology of battery measurements}
Battery characterisation methods measure the SoC and/or SoH of a cell and are necessary for understanding the charging process~\cite{Theuerkauf2022,Li2016} , in order to reduce wasted time during charging, and to investigate damaging effects such as overcharging or lithium plating \cite{Petzl2014}.
The simplest characterisation methods use measurements of cell voltage---either in open-circuit or during charge-discharge cycling---from which properties of the cell, such as operating range and capacity, can be deduced~\cite{Petzl2013}.
Here, open-circuit measurements are preferable, as they require less assembly time in industrial settings and electrical testing can introduce temperature rises, which need to be accounted for to take accurate measurements \cite{Schinagl2025}.

After charging or discharging, the voltage of a cell can remain transient for many hours during relaxation. 
Such relaxation processes are dependent on charge transfers, ionic-concentration distributions, and the ohmic resistance, and relaxation curves must be taken into account when estimating SoC in reduced time~\cite{Pei2013,Wang2024}.
In a typical measurement protocol, the inhomogeneous charge distribution reached during operation, with higher current densities located at the tabs, relaxes to a new homogeneous equilibrium in open-circuit configuration. 
We record the small electrical currents (current densities up to \SI{5.2}{\milli\ampere/\metre^2}) produced during this process using an array of sensitive magnetometers, enabling us to create an image of this activity across the cell. 
The method combines spatial and temporal resolution, providing direct access to internal dynamics within the cell---as well as localised features arising from battery geometry, production defects, or areas of increased ageing, which can be masked in voltage measurements.

\begin{figure}[hbt!]
\centering
\includegraphics[width=0.8\linewidth]{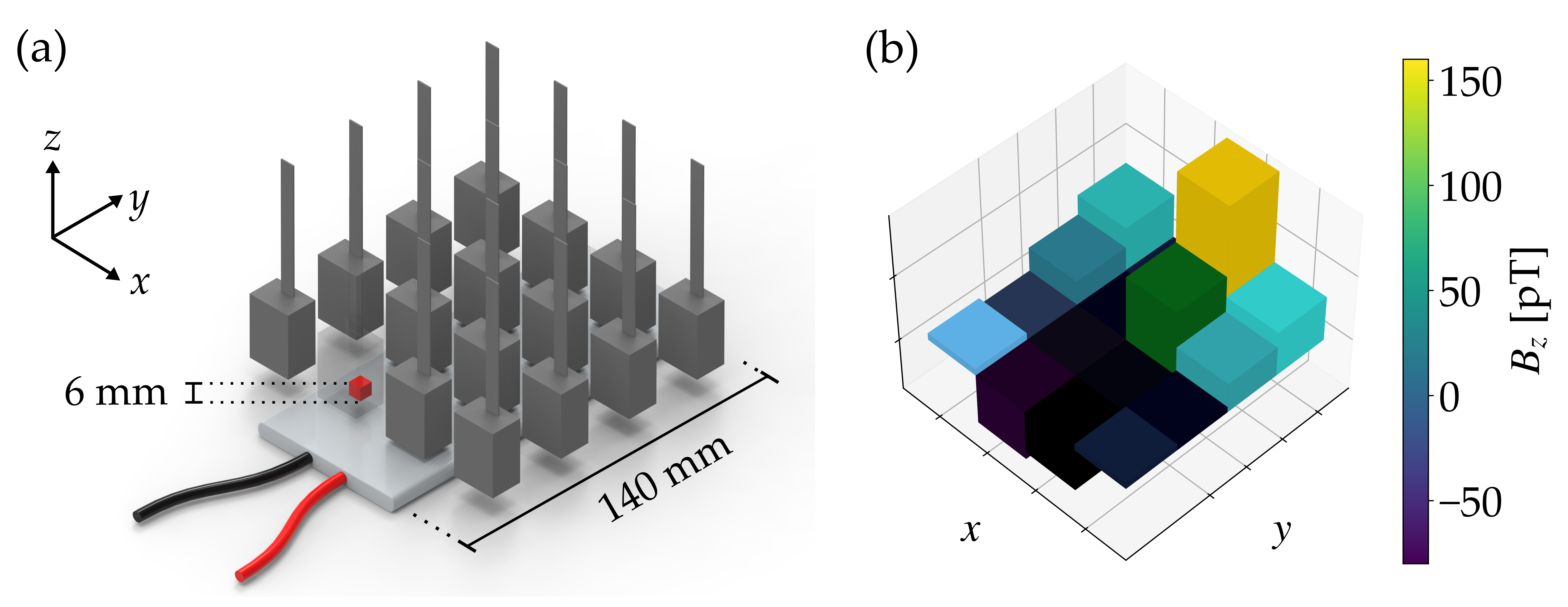}
\caption{: (a) General scheme for non-invasive, non-contact magnetic imaging of a battery using an array of optically pumped magnetometers (OPMs, grey boxes). The exact configuration of the array, including sensor density and orientation, can be adjusted depending on specific experimental requirements (Table~S1). Stand-off distance from the surface of the battery to the center of the atomic sensing volume (alkali vapour cell, depicted as a red cube) is less than 1\,cm. (b) Snapshot of a typical vector magnetic image of a battery, with each pixel corresponding to the reading (here, magnetic field along the $z$-axis) from a single sensor of the OPM array. This spatially resolved method allows us to distinguish the magnitude and sign of the field due to differing current densities at different locations within the battery at any given point in time.
}
\label{fig:Scheme}
\end{figure}

In our demonstration, we use a commercial flattened-jelly-roll cell (Powerstream PGEB-NM5858138-PCB) with dimensions \SI{138.5}{\milli\metre}$\times$\SI{58}{\milli\metre}$\times$\SI{6}{\milli\metre} and a nominal capacity of \SI{6000}{\milli\ampere\hour}. 
Further details as well as CT scans showing the current collectors can be found in the Supplementary Information (SI) and Fig.~S2.
In order to gain an initial estimate for the expected amplitude of current density relaxation signals, we constructed a three-dimensional finite element simulation ~\cite{Verbrugge2016,MarcDoyle1996,Karthikeyan2008,Thomas2002} for the simplest case of a single-layer cell.
Here, contributions to the current density from all regions reaches a maximum magnetic field of approximately \SI{700}{\femto\tesla} normal to the surface at a height of \SI{1}{\centi\metre} above the center of the cell and \SI{30}{\second} after discharge. 
The measured magnetic field strength can be enhanced from contributions of additional cell layers, while deviations from the predicted magnetic field may also arise due to the specific geometry of the roll structure which modifies the field distribution.
As we shall see, the actual magnetic field produced by the commercial multi-layer cell, measured by a sensor at a height of \SI{8}{\milli\metre}, can reach up to on order \SI{100}{\pico\tesla}.
As this signal evolves during relaxation, the measured field value eventually drops below the \SI{}{\pico\tesla}-range resolution of classical magnetometers.
Therefore, it is necessary to use more sensitive quantum magnetometers---specifically, OPMs or SQUIDs---in order to properly monitor the relaxation process. 
Although both quantum-magnetometer types are suitable for proof-of-principle measurements, the versatility and non-cryogenic operation of OPM systems make them the obvious choice for practical applications. 
A general scheme for magnetic imaging of battery current density using modular OPM arrays is depicted in Fig.~\ref{fig:Scheme}.

The OPMs used in this work were zero-field magnetometers (QuSpin QZFM Gen-2) with a magnetic-field sensitivity of $<$\SI{15}{\femto\tesla/\sqrthz} and a dynamic range of $\pm$\SI{5}{\nano\tesla}, with the ability to operate in ambient fields of up to \SI{50}{\nano\tesla}~\cite{Osborne2018}. 
The SQUID system used was a $304$-channel system mounted in the Berlin Magnetically Shielded Room (BMSR-2.1)~\cite{Voigt2013} with a magnetic-field sensitivity of \SI{3}{\femto\tesla/\sqrthz}, a dynamic range of $\pm$\SI{27}{\nano\tesla}, and the ability to operate in ambient fields of up to \SI{10}{\micro\tesla}.
While SQUID magnetometers offer certain benefits such as a larger frequency bandwidth, the need for cryogenic cooling increases demands on infrastructure---thereby precluding portable systems and increasing the stand-off distance to the sample of interest, due to the walls of the liquid-helium dewar. 
As mentioned, OPMs require no cryogenics, allowing for a more compact sensor system with less required maintenance and greater accessibility for new users. Critically, the reduced offset distance to the sample enables stronger signals to be measured with improved spatial resolution. These properties are essential for any real-world application of battery magnetic imaging, e.g. production-line quality control or operando SoH monitoring. Therefore, SQUID measurements are included here mainly for benchmarking purposes.
Although the most sensitive OPMs are zero-field magnetometers---such as those used in this proof-of-principle work---which must operate in an environment free of excess magnetic noise (like Earth's magnetic field), total-field unshielded or uncompensated devices with sub-pT resolution also exist and have begun to be commercialized. Thus, the requirement of magnetic shielding is not a fundamental one for OPM-enabled battery imaging.

Depending on the specific experiment, the OPM sensors were mounted above the battery cell in one of several configurations of a planar rectangular array.
Here, three distinct setups were used: a $4\times1$ array inside a table-top cylindrical magnetic shield constructed from three layers of mu-metal, a $2\times3$ array in the same shield, and a $4\times4$ array inside a magnetically shielded room (BMSR-2.1 at PTB Berlin). Further details of the magnetic shield and shielded room may be found in the SI.
In all setups, each sensor measured two components of the magnetic field at a single location. We use an orthogonal definition of the field components (Fig.~\ref{fig:Scheme})(a), with the $x$-direction across the width of the battery, the $y$-direction along the length of it away from the tabs, and the $z$-direction out of the plane of the cell. 
Descriptions of sensor placement and orientation for each setup are provided in Table~S1, with the corresponding schematics in Fig.~S1. Further experimental details about construction of the sensor holders are also included in the SI.
By contrast, the SQUID system was contained in a cryogenically cooled dewar, and the sensor used had a single-axis sensitivity along the $z$-direction and was located at a height of \SI{50}{\milli\metre} from the surface of the dewar. The necessary thermal insulation introduced a stand-off distance of \SI{50}{\milli\metre} from the sensor to the battery surface. 

\begin{figure}[hbt!]
\centering
\includegraphics[width=\linewidth]{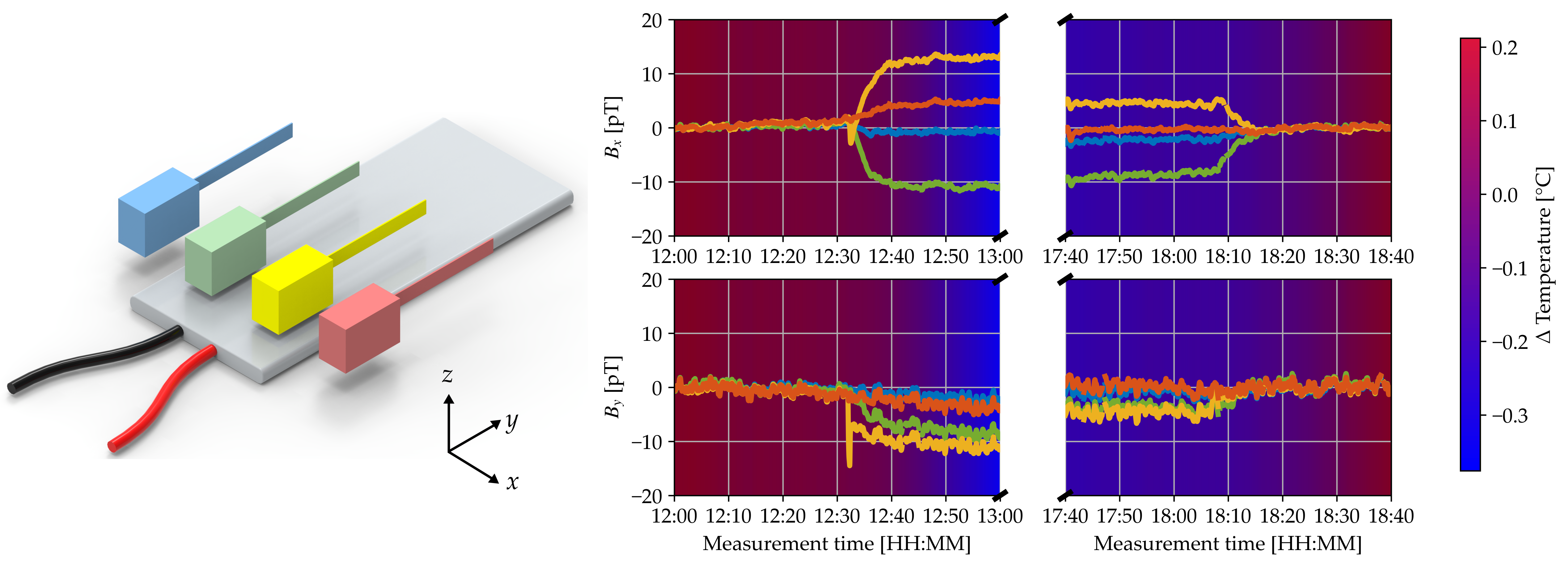}
\caption{: Time-series data from ambient measurements of a disconnected battery cell using a 1D $4\times1$ OPM array (top inset and Fig.~S1(a)), showing an example of a `jump' in the measured $B_x$ and $B_y$ magnetic fields within a period of \SI{1}{\hour}, out of a measurement over \SI{32}{\hour}. The background gradient depicts the cell temperature, as measured with a non-magnetic thermocouple mounted at the centre of the cell. Data were sampled at \SI{2}{\hertz}, with moving averages of \SI{20}{\second} and \SI{1}{\hour} used for magnetic field and temperature, respectively. See text for further details and discussion. Such data indicates that the OPM method is uniquely placed to detect localised phenomena of relatively small amplitude over longer timescales.}
\label{fig:Passive_measurement}
\end{figure}

The simplest possible way to showcase the long-term stability and spatial resolution of the OPM method is to conduct a `passive' measurement on a fully relaxed cell for many hours. An example of such a recording, taken using the one-dimensional array, is shown in Fig.~\ref{fig:Passive_measurement}.
During long-term open-circuit measurements of the disconnected cell (32\,h duration), we observed a series of step-like magnetic-field changes of approximately 10\,pT amplitude. These steps are distinct from abrupt field jumps caused by domain shifts in the mu-metal magnetic shielding~\cite{Barkhausen1919}, which occur within seconds, as the events here exhibit a gradual (10\,min) decay. The signal pattern---opposite-sign steps from sensors on the left and right flanks of the cell, respectively, which reverse after several hours, as well as a larger response from sensors positioned above the cell compared to those on the sides---indicates a transient, localised source of magnetic-field change within the cell. These events coincide with approximately \SI{0.1}{\degreeCelsius} temperature fluctuations recorded by a thermocouple positioned on the battery surface, on the centre line 23\,mm from the battery tabs. The coincident thermal and magnetic activity suggests a coupling to electrochemical or structural relaxation phenomena and/or local defects. Although infrequent and irregular, these observations represent direct evidence of dynamic, spatially dependent processes in the cell during extended open-circuit conditions, warranting further targeted studies.

In the remainder of this work, we focus instead on `active' magnetic-field measurements of induced relaxation processes, which are more controllable and therefore easier to interpret.
Open-circuit voltage relaxation is the process whereby the voltage reached by a battery cell during operation decays to a stable value---termed the open-circuit voltage---after being switched off. 
In EIS measurements, it is common to let a test cell rest for up to \SI{5}{\hour} for high-capacity (on the order of \SI{10}{\ampere\hour}) automotive batteries~\cite{Seitz2022}, with one study measuring relaxation times of \SI{24}{\hour}~\cite{Oldenburger2019}. 
Contemporary research has used voltage decays to understand different contributions to the total voltage relaxation from separate electrochemical components~\cite{Bernardi2011}. Spatially resolved measurements of magnetic-field relaxation over various timescales would allow for improved understanding of the changes in these contributions over the lifetime of a cell. The main purpose of this manuscript is therefore to elucidate the methodology for such measurements.
%----------------------- Experiment -----------------------

%---------------------- Relaxation ------------------------
\section*{Magnetic-field relaxation---Results}

\begin{figure}[hbt!]
\centering
\includegraphics[width=\linewidth]{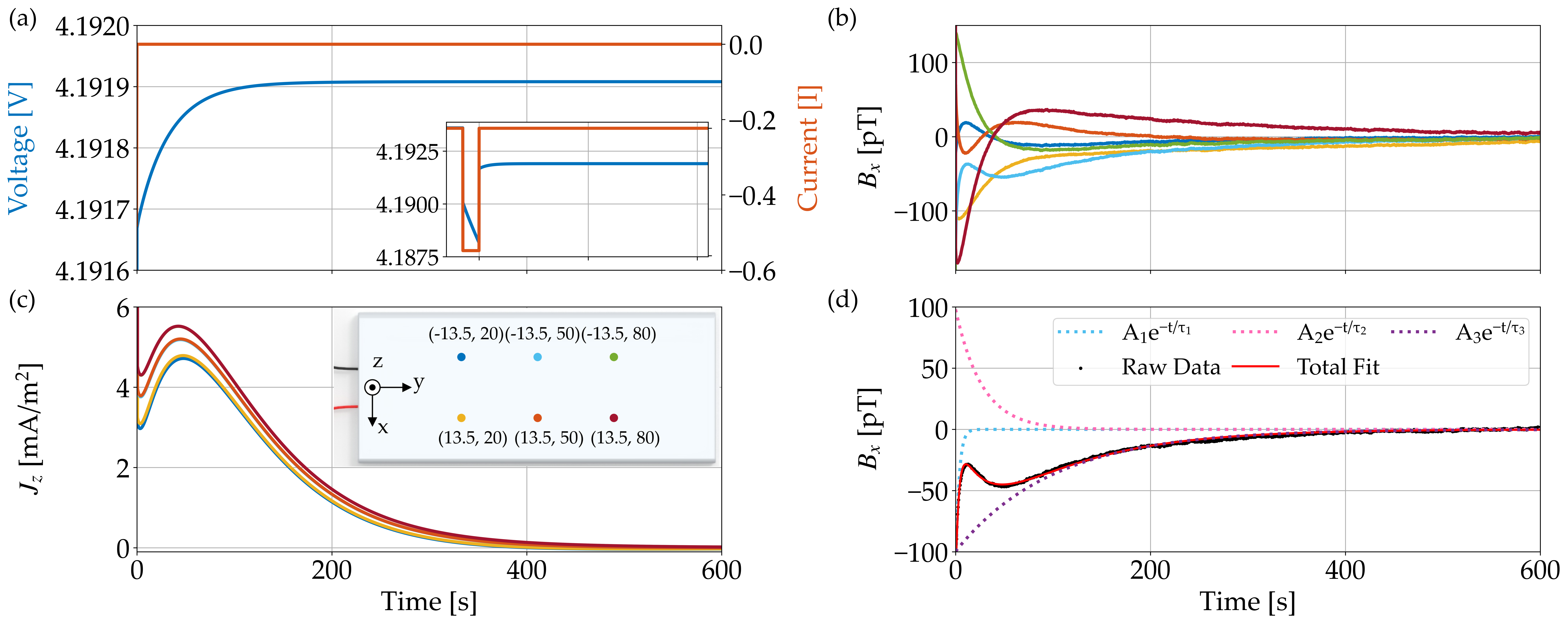}
\caption{:
(a) Example of relaxation in the cell voltage (blue) immediately after a single \SI{1}{\min}, \SI{0.5}{\ampere} (C/12) discharge pulse (orange), characterized by a relaxation time of \SI{33.21\pm0.02}{\second} at which the voltage decays to $1/e$ of its initial value.
The battery was fully charged and allowed to rest for \SI{4}{\hour} before the relaxation measurement was performed.
The inset shows the full voltage measurement during and after the discharge pulse, with ($x$,$y$) indicating sensor positions in mm.
(b) Measured magnetic field along the $x$-direction recorded with the 2D $2\times3$ OPM array (Fig.~S1(b)) immediately after the discharge pulse---see legend inset for colour-coding. 
The magnetic fields at each pair of sensors located the same distance from the cell tabs ($y$-coordinate) approximately mirror each other---indicating a symmetry of current density amplitude about the $y$-axis---with the largest amplitudes farthest from the tabs (green and red curves).
(c) Simulated electrolyte current density (see the section `Finite-element simulations') along the $z$-direction in the positive electrode, \SI{20}{\micro\metre} from the electrode-separator boundary of a \SI{62.4}{\milli\ampere\hour}-capacity single-layer cell following a discharge pulse.
The current density is shown at eight points in the electrode, at positions directly below the sensors in the OPM array. As in the magnetic-field data, panel (b), the two locations farthest from the tabs show the largest amplitude.
A peak in current density appears $\sim$\SI{30}{\second} after the pulse, followed by a decay from an average value of \SI{5.24 \pm 0.42}{\milli\ampere\per\square\metre} to zero on timescales characterized primarily by an `intermediate' relaxation time \SI{50.92}{\second} and a `slow' relaxation time \SI{66.43}{\second} (as extracted from a tri-exponential fit, see the section `Equivalent circuit models'), similar to the timescale of the magnetic-field response (b).
(d) Example relaxation measurement of a battery cell at 100\% SOC, recorded with an OPM at position (\SI{-13.5}{\milli\metre}, \SI{60}{\milli\metre}, \SI{8.2}{\milli\metre}) from the midpoint of the cell tabs.
Per Eq.~\eqref{eqn:ECM}, the sum of three exponential curves are fitted to the data---here with decay times $\tau_1 = \SI{4.6 \pm 2.0}{\second}$, $\tau_2 = \SI{20.3 \pm 3.5}{\second}$, and $\tau_3 = \SI{95.5 \pm 6.3}{\second}$---which gives an $R^2$-value of 0.994.
}
\label{fig:Relaxation}
\end{figure}

To induce relaxation in the battery, a \SI{0.5}{\ampere} square-wave discharge pulse was applied to the cell for \SI{1}{\min}.
This pulse corresponded to $0.14\%$ of the nominal capacity (\SI{6000}{\milli\ampere\hour}), so that the total SoC was largely unaffected throughout the measurement. 
The applied discharge pulse is shown in Fig.~\ref{fig:Relaxation}(a), alongside the measured voltage profile. 
It is possible to observe voltage relaxation: with the battery at \SI{4.19167}{\volt} immediately after the discharge pulse, the voltage recovers by +\SI{0.24}{\milli\volt} after $t=$~\SI{840}{\second}.
The corresponding magnetic-field components were measured using the two-dimensional $2\times3$ OPM array (Fig.~S1(b)), along the sensitive measurement axes $x$ and $y$, for \SI{14}{\min} (\SI{840}{\second}) following the pulse. The $x$-components of the magnetic field are presented in Fig.~\ref{fig:Relaxation}(b), where the static background field at $t=$~\SI{840}{\second} has been subtracted to isolate the relaxation signal. 
These results illustrate that the magnetic field varies across the battery, both in the initial amplitude of the field---between \SI{54}{\pico\tesla} and \SI{140}{\pico\tesla}---and in the qualitative shape of the decay. 
The relaxation signals include both monotonic decays and more complex dynamics that are best approximated with multiple exponential functions, as examined in the section `Equivalent circuit models'. 

\subsection*{Finite-element simulations}

To validate the experimental results and to provide greater insight into the underlying current density dynamics in the cell, we conducted a three-dimensional simulation using finite-element software. Complete details of the simulation software and package used can be found in the SI. 
We simplified the geometry of the multi-layer jelly roll cell to a single-layer rectangular cell of dimensions \SI{60}{\milli\metre}$\times$\SI{138}{\milli\metre}$\times$\SI{170}{\micro\metre} (L$\times$W$\times$H) and capacity \SI{62.4}{\milli\ampere\hour}.

In this simplification, behaviour along the length of the cell ($y$-direction) best matches the real-world setup, because the boundary conditions are similar at the tabs and at the side opposite the tabs. By contrast, the folding of layers of the real jelly roll cell means that magnetic field along the $x$-direction will have different behaviour from the simulation.
The applied C-rate during the simulation is C/12, as in the actual experiment, with the cell at 90\% SoC.
We found that the magnetic-field distribution in all three dimensions seconds after discharge-pulse switch-off reveals spatial variation of transient relaxation activity within the simulated cell (see images in Fig.~S3).
Furthermore, considering the spatial and time dependence of the $x$-component of the field, resultant from currents in the $y$- and $z$-directions, we obtained results qualitatively consistent with experiment: there is a change in direction of the magnetic field along the length of the cell, and this gradient evolves over the duration of the relaxation period (see images in Fig.~S4).
In Fig.~\ref{fig:Relaxation}(c), the spatially resolved time dependence of simulated current density $J_z$ is plotted for direct comparison with the measured $B_x$-field data. 
Extracting average relaxation times from the one-dimensional curves, according to the procedure outlined in the next section, reveals that timescales are similar for measured and simulated data, as expected (see caption of Fig.~\ref{fig:Relaxation}). 

\subsection*{Equivalent circuit models}

Equivalent circuit models (ECMs) are used as a tool for modelling the transient voltage dynamics of a battery cell.
One study~\cite{Bernardi2011} investigated relaxation effects attributed to individual domains of the cell, such as charge transfer in the positive electrode, using \SI{40}{\second} current pulses at 3\,C, 5\,C, and 10\,C in a \SI{200}{\milli\ampere\hour}-capacity nickel-cobalt-aluminium (NCA) cylindrical cell.
The authors identified multiple contributions to the voltage relaxation: an instantaneous drop caused by the ohmic resistance of the electrolyte and current collectors, a fast component (decay time well below one second) associated with charge-transfer processes at the electrode–electrolyte interface, and slower components (decay times greater than 20 s) attributed to concentration equilibration within the positive electrode. 
Another study \cite{Kindermann2015} specifically examined long-term relaxation behaviour and was able to distinguish liquid-phase concentration equilibration, with decay times of less than one minute, from solid-phase equilibration processes occurring over timescales of several hours. Similar models may be used for studying magnetic-field relaxation, where different contributions to the relaxation at various spatial locations can be isolated based on their associated decay times.

Combining three RC networks to form an ECM, the voltage drop over the networks corresponds to transient voltage behaviour.
For the case of magnetic-field relaxation, currents in the RC network are considered. 
Constructing a linear combination of the three main terms contributing to the total relaxation, we obtain for a time-dependent two-dimensional magnetic field
\begin{equation}
    B(x,y,t)=A_{1}e^{-t/\tau_1}+A_{2}e^{-t/\tau_2}+A_{3}e^{-t/\tau_3} \,.
    \label{eqn:ECM}
\end{equation}
Here, $\tau_i=R_iC_i$ ($i \in \{1,2,3\}$) is the characteristic decay constant of the RC network with resistance $R_i$ and capacitance $C_i$, and $A_i$ is a scaling factor quantifying the relative contribution of that network to the total magnetic field. Fig.~\ref{fig:Relaxation}(d) shows an example of fitting this model to experimental data recorded from a single sensor. Tri-exponential relaxation behaviour is clearly observed, and three relaxation timescales can be extracted corresponding to fast (on order 1\,s), intermediate (on order 10\,s), and slow (on order 100\,s) components. 
Additional exponential terms would increase the number of free parameters without adding meaningful interpretability and would risk over-fitting.
Interpretation of the underlying physical mechanisms which contribute to the three timescale behaviour is beyond the scope of this work, but we apply the fitting method to quantify our results and validate the magnetic-field measurements against voltage measurements and simulations. See Fig.~\ref{fig:Relaxation} for additional details.

\subsection*{Magnetic imaging}

\begin{figure}[hbt!]
\centering
\includegraphics[width=\linewidth]{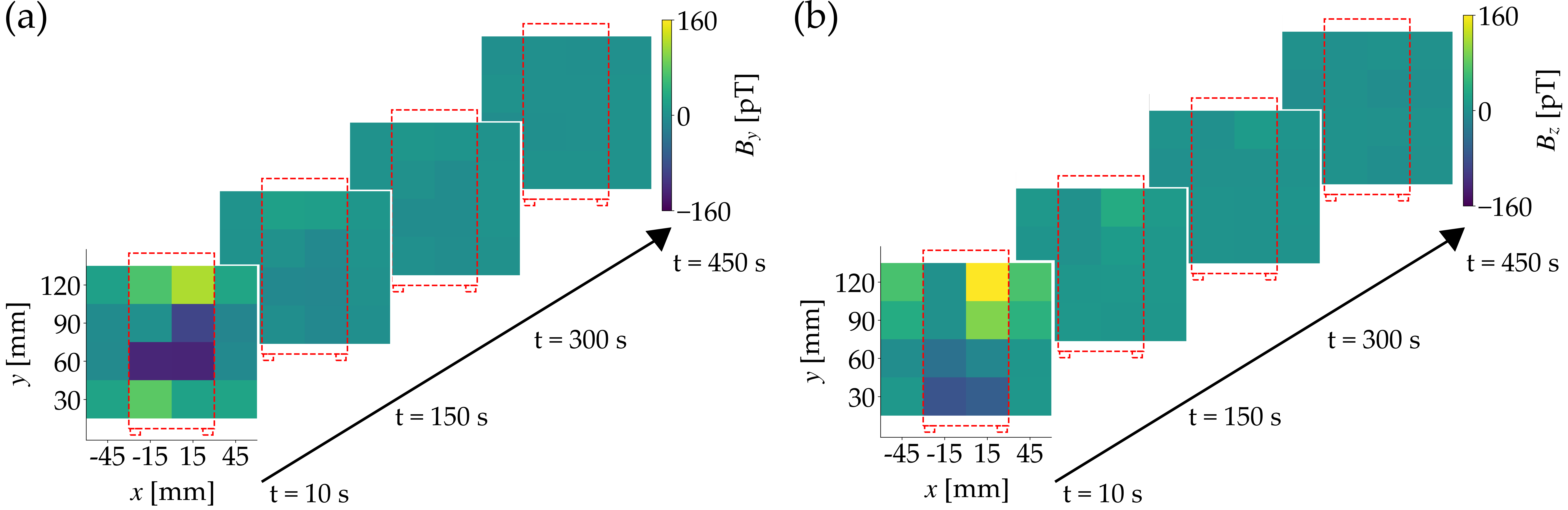}
\caption{: Time-resolved magnetic images of relaxation after a \SI{1}{\min} discharge of a fully charged cell with 0.1\,C (\SI{0.6}{\ampere}), collected with the 2D $4\times4$ OPM array, Fig.~\ref{fig:Scheme} and Fig.~S1(c). The images, recorded at times of \SI{10}{\second}, \SI{150}{\second}, \SI{300}{\second}, and \SI{450}{\second} following the discharge pulse, track evolution of (a) the $y$-magnetic field and (b) the $z$-magnetic field  as they decay to homogeneous equilibrium values. Red dashed lines demarcate the position of the cell beneath the sensors; the battery lies below the central $2\times4$ array of sensors, as indicated. Such images underscore the spatial and temporal variability which the OPM-based method is able to capture.
}
\label{fig:Images}
\end{figure}

It is frequently instructive to visualize the spatially resolved magnetic-field data from two-dimensional arrays as a pixelated heat map, where each pixel corresponds to the magnetic-field reading from a single sensor. Thus, a true magnetic image can be created at a given point in time, with the magnetometer array acting as a camera.
In Fig.~\ref{fig:Images}, magnetic images taken with the $4\times4$ OPM array, Fig.~S1(c), are presented as a time series over the 10\,min period following a 1\,min, 0.1\,C discharge pulse. 
The displayed changes in magnetic field along the $y$- and $z$-directions are calculated relative to the final equilibrium value at the end of discharge ($t =$~\SI{600}{\second}).
As is also borne out by simulations (Figs.~S3 and S4), the magnetic-field gradient is largest along the length of the battery ($y$-direction).

\subsection*{OPM and SQUID comparison}

\begin{figure}[hbt!]
\centering
\includegraphics[width=0.55\linewidth]{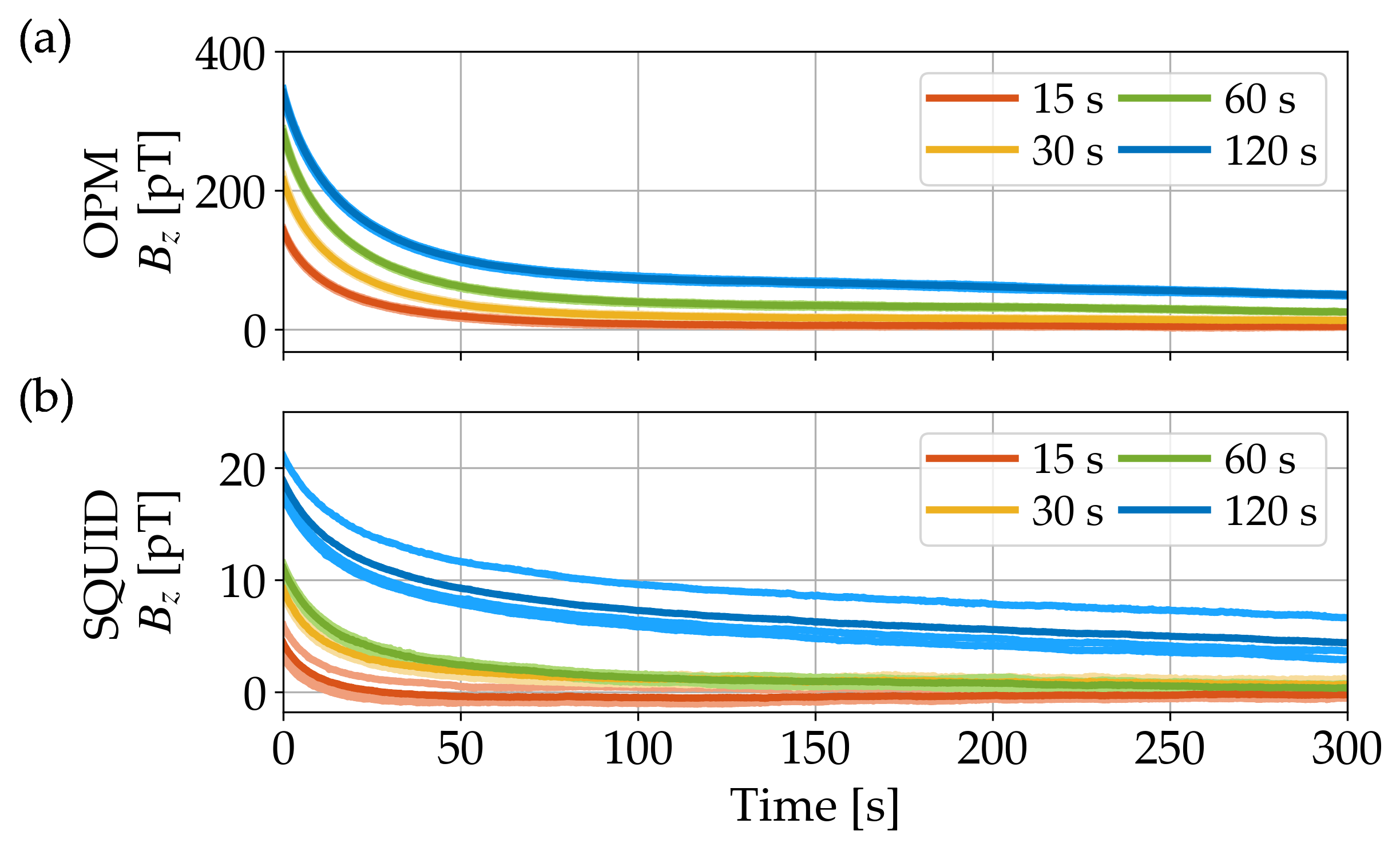}
\caption{: Comparison of equivalent OPM and SQUID recordings for method benchmarking and investigation of systematics.
(a) Results of an OPM measurement in the $z$-direction above a battery after \SI{0.6}{\ampere} (0.1\,C) discharge of durations \SI{15}{\second}, \SI{30}{\second}, \SI{60}{\second}, and \SI{120}{\second}. The corresponding average magnetic fields at time $t=0$ are \SI{142.8 \pm 3.0}{\pico\tesla}, \SI{215.2 \pm 5.9}{\pico\tesla}, \SI{283.6 \pm 4.1}{\pico\tesla}, and \SI{341.1 \pm 5.6}{\pico\tesla}. 
The experiment was conducted three times for each discharge duration, with the mean field value plotted as a darker curve.
Fitting monoexponential decays to the solid curves yields a characteristic average relaxation time of 20.5$\pm$0.9\,s, where the error bar corresponds to standard deviation.
(b) Results of an equivalent, simultaneous SQUID measurement recorded with the pick-up coil nearest to the OPM sensor and measuring the magnetic field along the same direction. 
The average fields at $t=0$ are \SI{4.4 \pm 1.2}{\pico\tesla}, \SI{9.2 \pm 0.3}{\pico\tesla}, \SI{11.0 \pm 0.4}{\pico\tesla}, and \SI{18.8 \pm 1.6}{\pico\tesla}.
The greater variance in the \SI{120}{\second} SQUID measurement (blue) is due to a systematic artefact in one trial, which has not been removed in order to keep the analysis of both datasets consistent.
Fitting monoexponential decays to the solid curves yields a characteristic average relaxation time of 18.3$\pm$5.0\,s, where the error bar corresponds to standard deviation.
}
\label{fig:SQUID_comparison}
\end{figure}

In order to benchmark the OPM-based method against another magnetic method, comparative measurements were taken with the OPM and SQUID systems. 
This also served as an investigation of possible systematic sensor effects, such as magnetic-field offsets and feedback loops.
For example, the OPM control system featured a PID-feedback loop on the active heater system of the atomic vapour cell to hold the vapour pressure inside the sensing volume constant, but this had to be turned off for the duration of measurement to prevent low-frequency oscillations as the OPM recovered from the large (\SI{10}{\micro\tesla}) pulse.
Similarly, the SQUID system required an offset compensation following the change in ambient field during discharge pulse.
Results from a single OPM and the closest equivalent SQUID sensor measuring along the same axis ($\hat{z}$) are compared in Fig.~\ref{fig:SQUID_comparison}. 
The comparative relaxation measurements were conducted using discharge currents of \SI{0.6}{\ampere} (0.1\,C) for durations of \SI{15}{\second}, \SI{30}{\second}, \SI{60}{\second}, and \SI{120}{\second}, where three different measurements were taken for each discharge duration to verify reproducibility.
Because the OPM had a stand-off distance of \SI{8.4}{\milli\metre}, whereas the SQUIDs had a stand-off distance of \SI{50}{\milli\metre}, the magnetic field measured by the OPM was 1--2 orders of magnitude larger. 
However, the respective relaxation curves display a similar shape and identical relaxation timescale---see Fig.~\ref{fig:SQUID_comparison} for further details---confirming that the method is reproducible for different sensor systems.
In both cases, it is possible to observe an increase in the initial amplitude of the magnetic field, and therefore in the current density distribution, with increasing discharge duration. 
Importantly, the run-to-run variation for a given duration is significantly (by at least an order of magnitude) less than the variation due to change in duration.

\subsection*{Dependence on discharge parameters and state of charge}

Having established the stability and reproducibility of the OPM-based method, a further detailed study was performed to investigate the influence of discharge parameters (amplitude and duration) and SoC on the magnetic-field relaxation. 

\begin{figure}[hbt!]
\centering
\includegraphics[width=\linewidth]{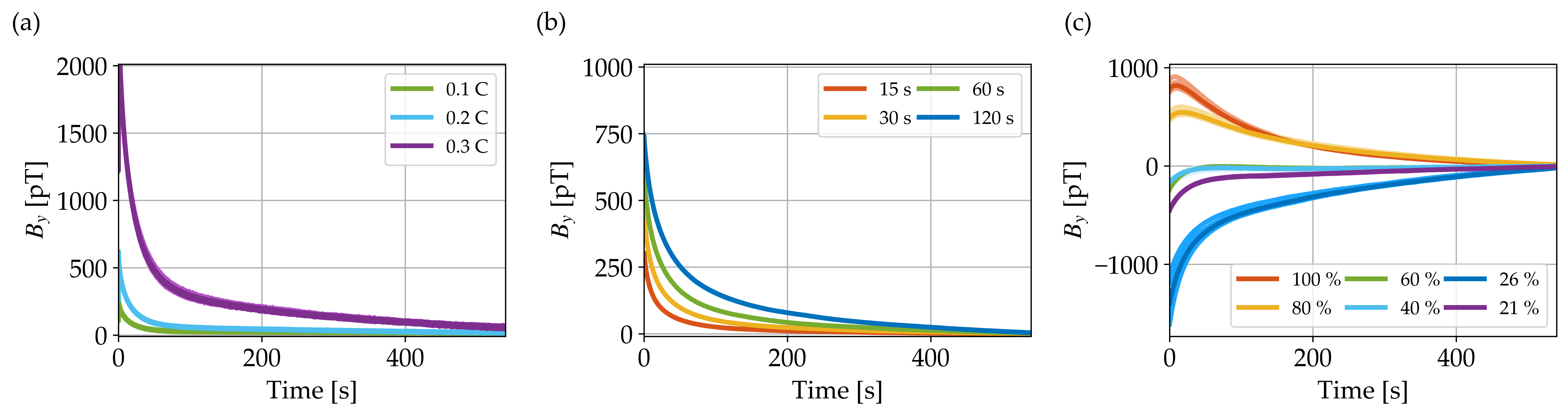}
\caption{: Measured dependence of magnetic-field relaxation along the $y$-direction on discharge parameters and state of charge. For a given set of parameters, the measurement was conducted three times, with the mean field value plotted as a solid curve and each measurement run shown in a faded curve. 
(a) Magnetic-field relaxation after a \SI{60}{\second} discharge pulse applied to the cell at 100\% SoC, with increasing currents: \SI{0.6}{\ampere} (0.1\,C), \SI{1.2}{\ampere} (0.2\,C), and \SI{1.8}{\ampere} (0.3\,C). Data are displayed from a single OPM within the 2D $4\times4$ array (Fig.~\ref{fig:Scheme} and Fig.~S1(c)) at position (\SI{-15}{\milli\metre}, \SI{100}{\milli\metre}, \SI{8.4}{\milli\metre}) from the midpoint of the tabs. 
(b) Similar OPM measurements for increasing duration of a \SI{0.6}{\ampere} (0.1\,C) discharge pulse applied to the cell at 100\% SoC: \SI{15}{\second}, \SI{30}{\second}, \SI{60}{\second}, and \SI{120}{\second}.
(c) Magnetic-field relaxation recorded while decreasing the SoC from 100\% to 21\%. Data are displayed from a single OPM within the 2D $4\times4$ array at position (\SI{-15}{\milli\metre}, \SI{90}{\milli\metre}, \SI{8.4}{\milli\metre}) from the midpoint of the tabs. The shape of the relaxation curve evolves from a peaked decay at 100\% SoC to a monotonic decay at 60\% SoC and below, with the initial magnetic field also changing sign---indicating that magnetic imaging is a feasible and promising tool for SoC estimation.
}
\label{fig:Control_parameters}
\end{figure}

First, the applied discharge current was stepped from \SI{0.6}{\ampere} (0.1\,C) to \SI{1.2}{\ampere} (0.2\,C) to \SI{1.8}{\ampere} (0.3\,C), while the duration was maintained at \SI{60}{\second}. 
The measured magnetic field in the $y$-direction is shown in Fig.~\ref{fig:Control_parameters}(a).
Then, a constant discharge current of \SI{0.6}{\ampere} (0.1\,C) was applied to the cell, while the duration of the current was stepped from \SI{15}{\second} to \SI{30}{\second} to \SI{60}{\second} to \SI{120}{\second}. 
The measured magnetic field in the $y$-direction is shown in Fig.~\ref{fig:Control_parameters}(b).
For both of these experiments, results are displayed as the average of three recordings.

Corresponding 1D simulations of the current density inside the cathode of a single-layer cell are presented in Figs.~S5(a) and S5(b). 
To facilitate validation with reduced computational cost, a simplified one-dimensional version of the three-dimensional simulation was employed.
In both experiments and simulations, the amplitude of magnetic field or current density increased with increasing discharge current and increasing duration, as expected.
However, the different shapes of the measured and simulated relaxation profiles indicate that the one-dimensional model is limited for approximating entire realistic cells, implying that geometric effects are significant when considering electrochemical processes.

The feasibility of using magnetic imaging as a tool to improve SoC estimation is of particular interest, as this is an application where open-circuit voltage relaxation is used to improve measurement accuracy \cite{Wang2024}.
Experimental results from a single sensor at position (\SI{-15}{\milli\metre}, \SI{90}{\milli\metre}, \SI{8.4}{\milli\metre}) in the $4\times4$ array (Fig.~\ref{fig:Scheme} and Fig.~S1(c)), while decreasing the SoC from 100\% to 21\%, are plotted in Fig.~\ref{fig:Control_parameters}(c).
Here, the battery was fully charged to its maximum operating voltage of \SI{4.2}{\volt} at C/10 (\SI{0.6}{\ampere}), then left to rest for \SI{12}{\hour} before the first measurement. 
Between each measurement, it was discharged sequentially in steps of \SI{1.2}{\ampere\hour}, C/5, decreasing the SoC from $100\%$ to $21\%$ with three recordings at each SoC.
An additional reading was taken at $26\%$ after the battery reached a cut-off voltage of \SI{3.0}{\volt} during the fourth discharge.
The measured total discharge capacity of the cell during this experiment was \SI{4753}{\milli\ampere\hour}, in comparison to the nominal \SI{6000}{\milli\ampere\hour}.
As seen in the Fig.~\ref{fig:Control_parameters}(c), the shape of the decay curve changes dramatically and reproducibly as a function of SoC.

Once again, we compared our experimental data with the results of a one-dimensional simulation, plotted in Fig.~S5(c). 
The qualitative differences between the measured and simulated SoC-dependent relaxation curves suggest that more detailed simulations are required to model the full electrochemical dynamics being observed.

From an experimental perspective, in all three studies shown in Fig.~\ref{fig:Control_parameters}, the consistency between runs with parameters held constant indicates that the results are reproducible. 
As in the OPM-SQUID benchmarking study (Fig.~\ref{fig:SQUID_comparison}), the run-to-run variation in measured magnetic field is significantly less than the variation due to a parameter change, demonstrating the robustness of the method for critical real-world applications such as SoC estimation.

\subsection*{Extraction of relaxation parameters and comparison to EIS}

\begin{figure}[hbt!]
\centering
\includegraphics[width=\linewidth]{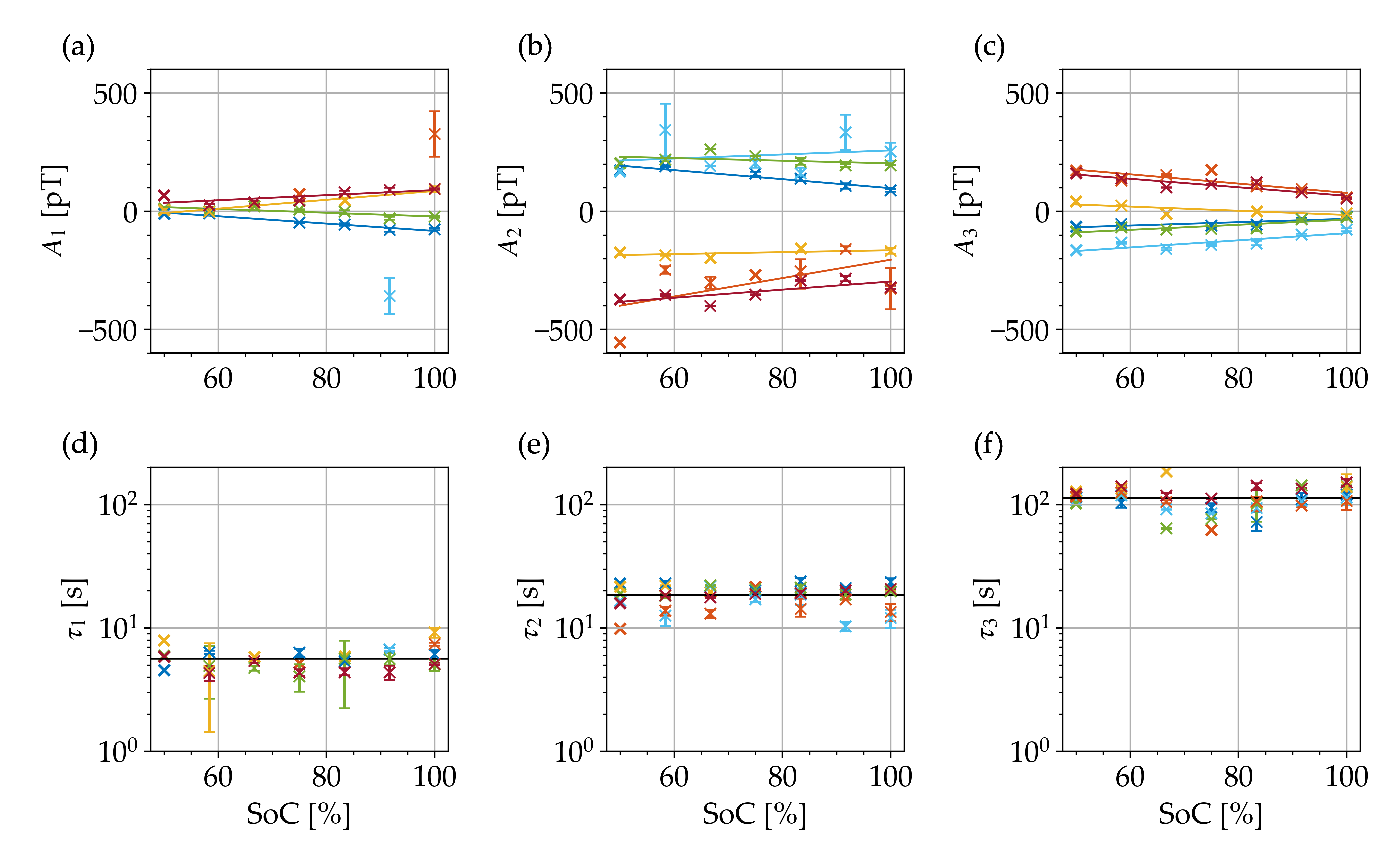}
\caption{:
Relaxation parameters extracted from measurements conducted on the sample cell with SoC between $50\%$ and $100\%$ (\SI{3000}{\milli\ampere\hour} and \SI{6000}{\milli\ampere\hour}, respectively), after a discharge pulse at C\,/12. The recordings were made with the $2\times3$ OPM array depicted Fig.~S1(b), with the 100\% SoC recording corresponding to the time-resolved images in Fig.~\ref{fig:Images}(b). A tri-exponential function was fitted to the magnetic-field data for each SoC (as discussed in the SI), according to the equivalent circuit model, Eq.~\eqref{eqn:ECM}.
For each of the six locations in the array, the extracted amplitudes $A_i$---panels (a,\,b,\,c)---and the decay times $\tau_i$---panels (d,\,e,\,f)---are plotted for the three terms of the fit function. We see that these three terms correspond to three different relaxation timescales: `fast', panels (a--d), `intermediate', panels (b--e), and `slow', panels (c--f). The intermediate contribution appears to dominate most of the readings at all measured SoC. See text for further discussion and interpretation.
Error bars on the extracted parameters correspond to the minimum–maximum variation obtained from repeated measurements and fitting procedures; markers without error bars correspond to a single dataset.
Details of the fitting procedure, as well as the criteria used to remove outliers, are provided in the SI. Lines on panels (a,\,b,\,c) correspond to linear trends, as a visual aid. The black lines on panels (d,\,e,\,f) correspond to the average timescales over all positions and SoC.
}
\label{fig:Relaxation_parameters}
\end{figure}

To demonstrate the full qualitative and quantitative power of our method for magnetic imaging of battery relaxation, Eq.~\eqref{eqn:ECM} was applied to investigate spatial dependence of field decay as a function of SoC after pulsed discharge at 1/12\,C. 
For this study, the data were collected using the $2\times3$ OPM array (Fig.~S1(b)), with the SoC stepped between 50\% and 100\%. 
All extracted relaxation parameters from the $B_x$ magnetic data---characterizing three different contributions to the overall relaxation---are displayed in Fig.~\ref{fig:Relaxation_parameters}. 
The fitting procedure was introduced in Fig.~\ref{fig:Relaxation}(d) and the surrounding discussion.
For all six sensors in the array, the amplitudes $A_i$ and the relaxation times $\tau_i$ are extracted for each of the three terms of the fit function, Eq.~\eqref{eqn:ECM}, and plotted against the SoC. 
It can be seen that each term has a distinct timescale, spanning fast (on order 1\,s), intermediate (on order 10\,s), and slow (on order 100\,s) decay.

Although the focus of this work is on developing a methodology rather than extracting physical or chemical information about specific batteries or batches, several features of the results in Fig.~\ref{fig:Relaxation_parameters} are striking and worthy of comment. 
First, the relaxation times $\tau_i$ (Fig.~\ref{fig:Relaxation_parameters}(d)--(f)) appear independent of both SoC and spatial location, Fig.~\ref{fig:Relaxation_parameters}(d). Second, the amplitude $A_2$ of the intermediate decay term is clearly dominant and also largely independent of SoC for a given spatial location (Fig.~\ref{fig:Relaxation_parameters}(b)), but spatial dependence of the amplitude is observed for all SoC.
Third, the amplitudes $A_1$ (Fig.~\ref{fig:Relaxation_parameters}(a)) and $A_3$ (Fig.~\ref{fig:Relaxation_parameters}(c)) show increasing and decreasing trends, respectively, with increasing SoC.
% Second, while the global or average value of $\tau_3$ also appears approximately independent of SoC, Fig.~\ref{fig:Relaxation_parameters}(f), the spatial variation increases from \SI{115.5 \pm 9.3}{\second} at 50\% SoC to \SI{128.7 \pm 17.5}{\second} at 100\% SoC.
% The associated amplitude $A_3$, Fig.~\ref{fig:Relaxation_parameters}(e), decreases with increasing SoC.
Fourth, not all positions and states of charge exhibit all three exponential components---this is most clearly seen in the fast relaxation component, which is absent for some datasets.
We also note that the majority of data points for the central two positions in the $2\times3$ array are excluded for visualisation due to insignificant improvements in the fit quality (see SI for more details on the fitting procedure and plotting exclusions).
Fifth, a common property can be observed for all extracted amplitudes $A_i$: namely, the initial magnetic fields measured on opposite sides of the battery--represented by `warm' colours yellow, orange, and red, versus `cool' colours blue, cyan, and green (see legend of Fig.~\ref{fig:Relaxation})---display opposite sign for all SoC.
All of these observations warrant further targeted studies to correlate relaxation data with internal processes in the cell; it is also reasonable to expect that more than three underlying relaxation mechanisms exist that group broadly into three different timescales. 
These results demonstrate that quantum magnetic imaging, in contrast to less sensitive global methods, reveals rich spatially dependent behaviour during battery relaxation.

\begin{figure}[hbt!]
\centering
\includegraphics[width=\linewidth]{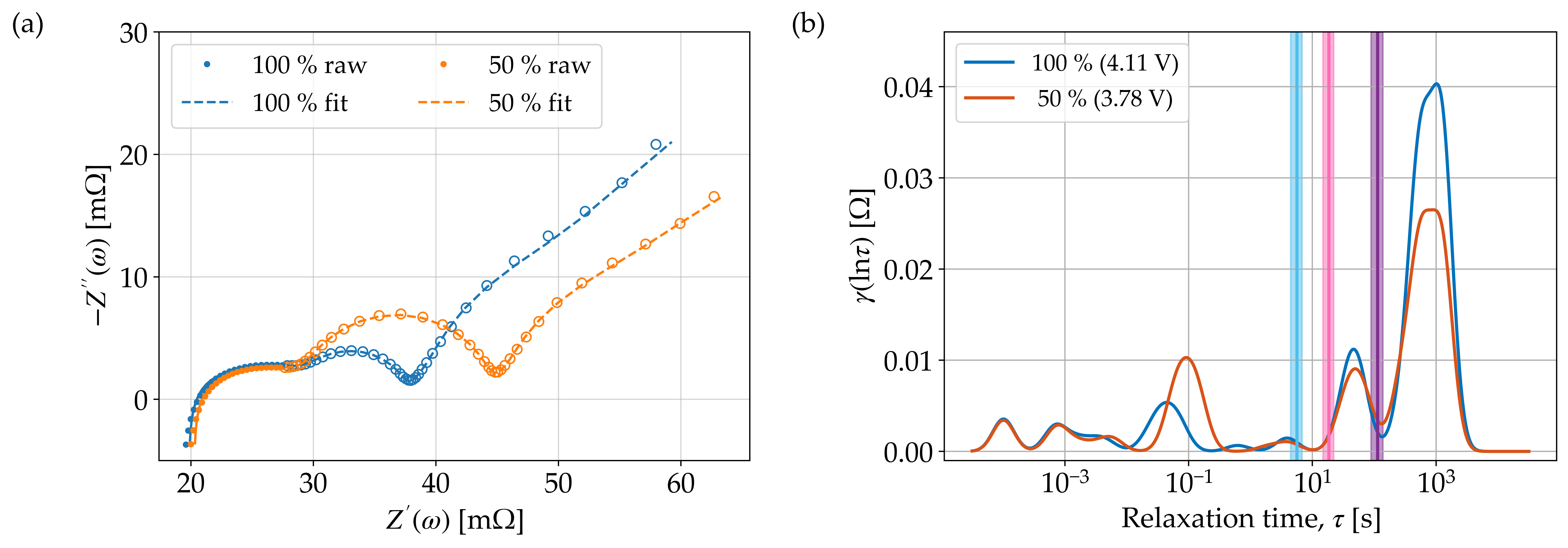}
\caption{:
(a) Nyquist plot showing the impedance of the test cell at $100\%$ SoC (\SI{4.11}{\volt}, blue) and $50\%$ SoC (\SI{3.78}{\volt}, orange) recorded using EIS with 85 frequencies between \SI{0.8}{\milli\Hz} and \SI{6}{\mega\Hz}. The measured response is shown as data points and fits (solid curves) are overlaid.
(b) Plot of the distribution of relaxation times---corresponding to the data set in panel (a) and used to construct the fits there---revealing three main relaxation times as peaks in the distribution. For comparison, the averaged relaxation times extracted from all magnetic measurements (Fig.~\ref{fig:Relaxation_parameters}) are plotted as vertical lines with shaded areas showing the standard deviations with cyan, pink and purple showing the fast, intermediate and slow timescales respectively. See text for further details and discussion.}
\label{fig:EIS_comparison}
\end{figure}

Finally, we directly compared our OPM measurements with EIS measurements of the same battery cell. 
While EIS is a global and not a spatially resolved method, it can also be employed to extract decay times from relaxation data. 
Fig.~\ref{fig:EIS_comparison}(a) shows the EIS data presented in a Nyquist plot \cite{Murbach2020}. 
Two EIS measurements, with the cell at $100\%$ SoC (\SI{4.11}{\volt}) and $50\%$ SoC (\SI{3.78}{\volt}), were taken using a Zahner Zennium XC device configured to measure at 85 frequencies between \SI{0.8}{\milli\Hz} and \SI{6}{\mega\Hz} in potentiostatic mode with \SI{5}{\milli\volt} amplitude. 
In both measurements, the cell was fully charged at C/6 (\SI{1.0}{\ampere}) until a current of \SI{0.6}{\ampere} (C/10) was reached, allowed to relax for \SI{24}{\hour}, then discharged with a current of \SI{2}{\ampere} (C/3) for \SI{1.5}{\hour}, followed by a period of rest for \SI{2}{\hour}. 
Throughout the experiment, the cell surface temperature was held at \SI{23}{\degreeCelsius}. 
In the Nyquist plot, the real part of the impedance, $Z^{\prime}\left(\omega\right)$ is displayed against the negative imaginary part, $-Z^{\prime\prime}\left(\omega\right)$. 
As both quantities are frequency-dependent, each data point corresponds to a specific frequency, with lower frequencies appearing farther to the right where the difference between the two curves is greatest. 

In order to calculate the distribution of relaxation times (DRT) from a Nyquist plot and highlight the most significant contributions to the EIS spectrum, the ECM may be generalised to comprise of an infinite number of terms~\cite{Wan2015}. 
The resulting DRT plot displays the distribution function $\gamma\left(\ln{\tau}\right)$ of relaxation times $\tau$, providing a relative measure of the density of relaxation processes occurring at different timescales.  
Fig.~\ref{fig:EIS_comparison}(b) shows the DRT corresponding to the Nyquist plots in Fig.~\ref{fig:EIS_comparison}(a).
For $50\%$ ($100\%$) SoC, the DRT plots exhibit three dominant peaks at relaxation times of approximately  $\tau$ = \SI{90}{\milli\second} (\SI{44}{\milli\second}), \SI{49}{\second} (\SI{47}{\second}) and \SI{900}{\second} (\SI{1000}{\second}).
Based on comparison with literature \cite{Zhao2022a, Zhao2022b} the first peak is assigned to charge-transfer relaxation, the second to liquid-phase relaxation, and the third to the solid-phase relaxation in the electrodes. The latter peak likely represents an overlap of two partially resolved contributions, corresponding to the solid-phase relaxation of the anode and cathode, respectively.
For comparison, the relaxation times extracted from the OPM measurements (Fig.~\ref{fig:Relaxation_parameters}) are overlaid as vertical lines, the distribution of which corresponds to the spatial variation recorded by individual sensors.
Temporal overlap can be observed between the intermediate EIS relaxation peaks around 50\,s and the bands corresponding to the OPM-detected relaxation times $\tau_2$ and $\tau_3$. 
Future studies will enable direct comparison between the two time-resolved measurements under identical environmental conditions. 
Here, the cell surface temperature was on average \SI{36}{\degreeCelsius} for the magnetic measurements and \SI{23}{\degreeCelsius} for the EIS measurements, which could affect relaxation times in the battery cell.
Even so, is important to keep in mind that the two methods may actually be sensitive to different underlying relaxation mechanisms. 
This comparative approach provides an interesting intersection of industrial methods and magnetic measurements, ripe for exploration in dedicated studies.
Although the EIS method can more easily capture high-frequency processes---due to the limited (on order 100\,Hz) bandwidth of the OPM sensors---acquisition times are typically much longer, with collection of a single Nyquist plot often taking hours.
These measurements also include mHz frequency measurements, which are omitted in these magnetic measurements due to the drift of the magnetic environment and of the sensors at similar time scales, which could be optimised in future work.

Single EIS measurements are less informative for SoC estimation and require careful characterisation for each cell, due to similarities in the features of EIS response curves at different SoC.
This underscores the need for a diverse set of analytical methods in the dynamic processes of lithium-ion cells with different SoC conditions, where equivalent circuit models can be used to investigate the various relaxation processes. 
Magnetic imaging with OPMs is just such a complementary tool, providing spatially resolved information in real time. 
%---------------------- Relaxation -----------------------

%---------------------- Conclusion -----------------------
\section*{Conclusions and outlook}
We have presented a novel technique for spatially resolving the relaxation of current density in a lithium-ion battery cell after pulsed discharge, using magnetic-field imaging. 
This proof-of-principle demonstration employed quantum sensors---both OPMs and SQUIDs---to measure magnetic fields in the \SI{}{\pico\tesla}-range and below, after a period of discharge with currents up to 0.3\,C. 
We also demonstrated the ability of the technique to capture rare/transient spontaneous activity within the cell, which warrants further investigation. 
In addition, sensitivity to variations in discharge parameters and SoC was confirmed, with SoC estimation emerging as a particularly promising practical application of our method.

The magnetometry results were compared with both one-dimensional and three-dimensional simulations, and analysis of the field relaxation using equivalent circuit models was carried out to quantitatively evaluate the relaxation process at multiple timescales.
In parallel, we conducted electrochemical-impedance-spectroscopy measurements of the same cell, comparing the ability of both methods discriminate to timescales of dynamic internal processes. 
Continued research is required to fully investigate the advantages of spatially localised impedance measurements over existing global frequency-response techniques. 
Understanding relaxation dynamics in different regions of the cell can contribute to the optimisation of battery design, performance, and longevity.
While this work focused on development of a method and did not attempt to extract specific information about underlying battery processes, observed spatial differences in magnetic behaviour already suggest varying current densities or ion-diffusion rates across regions.

Magnetic imaging lends itself to battery diagnostics through non-contact measurement of the current density during operation, location of defects and inhomogeneities within the cell, and tracking of ageing mechanisms.
The technique can also be employed in tandem with electrochemical modelling of cells for visualisation of current density patterns and underlying dynamics.
Development of new battery-specific OPM arrays with improved spatial resolution and higher bandwidth will allow for more thorough investigation of the relaxation process and comparison to high-frequency EIS measurements. 
Deployment of total-field OPM variants, which can operate unshielded in Earth's magnetic field or in the presence of magnetic materials, will also expand the practical reach of the method. 
Because magnetic imaging can be used for spatially resolved detection of diverse current density sources, including but not limited to Li-ion batteries, it is likely to form a key part of the battery-characterisation toolkit in the years to come.

%---------------------- Conclusion -----------------------

%---------------------- Additional -----------------------
\section*{Acknowledgements}
The authors acknowledge support from the Quantum Valley Lower Saxony iLabs-QBatt-Project (03ZU1209LC) and High-Tech Incubator (QVLS-HTI) program, as well as Lower Saxony's Ministries of Science and Culture and of Economic Affairs, Transport, Building and Digitalisation. This work was supported by the UK Quantum Technologies Hub for Sensors and Timing (EPSRC Grant EP/T001046/1), the University of Sussex Strategic Development Fund, and Innovate UK: Batteries - ISCF 42186 Quantum sensors for end-of-line battery testing. Funding was also provided by the Federal Ministry of Education and Research as part of the research cluster “Batterienutzungskonzepte” through the project MADAM4Life (grant number 03XP0327C).

We acknowledge the support of J. Voigt, S. Knappe-Gr\"uneberg, R. K\"orber, and T. Sander-Th\"ommes in conducting the magnetic-field measurements using the BMSR-2.1. 
Furthermore, we thank G. Stahl for technical assistance in collecting the electrochemical-impedance-spectroscopy data and M. Bason for his insightful discussions surrounding the simulations.

\section*{Author contributions}
WE, TC, MS, DW and PK conceived and designed the experiments. WE, TC, MS and DW performed the experiments. WE, TC, MW, AMF, GDK, MS, DW, PK and DUS analysed the data. 
WE, TC, MS and DW contributed materials and/or analysis tools. WE, TC, MW, AMF, GDK, PK and FO wrote the manuscript with input from all authors.

\section*{Competing interests}
The use of optically pumped magnetometers to determine the condition of an electrical charge storage device in a passive state is covered by patent GB2586829 held by CDO² Ltd and the University of Sussex. Other patents are pending.

\section*{Data availability}
The datasets generated during and/or analysed during the current study are available from the corresponding author on reasonable request.
%---------------------- Additional -----------------------

%---------------------- Bibliography ----------------------
\bibliography{BatteryDischarge}
%---------------------- Bibliography ----------------------

\end{document}